\documentclass[12 points]{article}
\usepackage{amssymb}
\usepackage{eucal}
\usepackage{amsmath}
\usepackage{amsthm}
\usepackage{graphicx}
\usepackage{float}
\usepackage{framed}

\newcommand{\R}{{\mathbb  R}}
\newcommand{\wt}{\widetilde}
  \numberwithin{equation}{section} 
  
  \theoremstyle{remark}

\begin{document}

\title{\large\bf Newton Algorithm on Constraint Manifolds and the 5-electron Thomson problem}
 \author{Petre Birtea, Dan Com\u{a}nescu\\ {\small Department of Mathematics, West University of Timi\c soara}\\ {\small Bd. V.
P\^ arvan,
No 4, 300223 Timi\c soara, Rom\^ania}\\ {\small birtea@math.uvt.ro, comanescu@math.uvt.ro}} 

\date{ } \maketitle

\begin{abstract}
We give a description of numerical Newton algorithm on a constraint manifold using only the ambient coordinates (usually Euclidean coordinates) and the geometry of the constraint manifold. We apply the numerical Newton algorithm on a sphere in order to find the critical configurations of the 5-electron Thomson problem. As a result, we find a new critical configuration of a regular pentagonal type. We also make an analytical study of the critical configurations found previously and determine their nature using Morse-Bott theory. Last section contains an analytical study of critical configurations for Riesz s-energy of 5-electron on a sphere and their bifurcation behavior is pointed out. 

\end{abstract}



\section{Introduction}

 Newton algorithm is a efficient numerical iterative method that can be applied for finding the critical points of the cost function. This type of algorithm uses second-order information about the cost function, fact that guarantees super-linear convergence (in cases it converges). Initially it was constructed on vector spaces, but later was adapted to manifolds for practical reasons, see \cite{absil-book} for a careful and detailed discussion of the geometry involved in such a context. 

Let $(S,{\bf g}_{_S})$ be a smooth Riemannian manifold and $\wt{G}:S\rightarrow \R$ be a smooth cost function. The iterative scheme is given by, see \cite{absil-book}:
\begin{equation}
x_{n+1}=\wt{\mathcal{R}}_{x_n}(\wt{{\bf v}}_{x_n}),
\end{equation}
where $\wt{\mathcal{R}}:TS\rightarrow S$ is a smooth retraction and the tangent vector $\wt{{\bf v}}_{x_n}\in T_{x_n}S$ is the solution of the {\bf (contravariant) Newton equation}
\begin{equation}\label{Newton-equation-standard}
\mathcal{H}^{\wt{G}}(x_n)\cdot \wt{{\bf v}}_{x_n}=-\nabla_{{\bf g}_{_S}}\wt{G}(x_n).
\end{equation}
In the above equation we have the Hessian operator $\mathcal{H}^{\wt{G}}:TS\rightarrow TS$. The link between the Hessian operator $\mathcal{H}^{\wt{G}}$ and the associated symmetric bilinear form $\text{Hess}\,{\wt{G}}:TS\times TS\rightarrow \R$ is given by:
$$\mathcal{H}^{\wt{G}}(x)\cdot \wt{{\bf v}}_x={\bf g}_{_S}^{\sharp}(x)\left(\text{Hess}\, {\wt{G}}(x)\cdot \wt{{\bf v}}_x\right),$$
where ${\bf g}_{_S}^{\sharp}(x):T^*_xS\rightarrow T_xS$ is the sharp operator in Riemannian geometry. The (contravariant) Newton equation \eqref{Newton-equation-standard} can be rewritten as
$ \text{Hess}\, {\wt{G}}(x_n)\cdot \wt{{\bf v}}_{x_n}=-{\bf g}_{_S}^{\flat}(x_n)\cdot \nabla_{{\bf g}_{_S}}\wt{G}(x_n)$,
or equivalently as the {\bf (covariant) Newton equation}: 
\begin{equation}\label{Newton-equation-forms}
 \text{Hess}\, {\wt{G}}(x_n)\cdot \wt{{\bf v}}_{x_n}=-d\wt{G}(x_n).
\end{equation}
To solve the above equation on the manifold $S$ one needs to put local charts on the manifold in order to make the computations and to have a good knowledge on the Riemannian geometry of the manifold. Another way is to embed the manifold $S$ in a larger manifold (usually an Euclidean space) and work with local charts (Euclidean coordinates) on this ambient space. The details of these constructions in the case when $S$ is described by constraint functions are given in the Section 2. We end this section presenting a new form of the Newton algorithm on constraint manifolds that we call {\bf Embedded Newton algorithm}.
The construction relies on the formula, given in \cite{Birtea-Comanescu-Hessian}, for the Hessian of a cost function restricted to a manifold defined by constraint functions, formula which involves only ambient coordinates.

In Section 3 we apply the Embedded Newton algorithm to the 5-electron Thomson problem. We find numerically the following critical configurations: bi-pyramidal configuration, square right pyramid configuration and a new family of saddle critical points in the configuration of a regular pentagon. 
We also carry out the analytical study of the three critical configurations discovered numerically. Each type of critical configuration comes as differential curves of critical points for the {\bf embedded gradient vector field} introduced in \cite{birtea-comanescu} and \cite{Birtea-Comanescu-Hessian}.  We prove that all this three critical configurations are non-degenerate in the sense of Morse-Bott theory.

In Section \ref{Riesz} we proceed with an analytical study of the Riesz s-energy of 5-points on a sphere problem which is a generalization of 5-electron Thomson problem. First we prove that the three types of critical configurations of the 5-electron Thomson problem discussed in Section 3 remain critical configurations for the Riesz s-energy problem. Various bifurcation phenomena do appear when $s$ varies and we find a new bifurcating value for $s=13.5204990011...$ when the square right pyramid goes from a saddle critical point into a local minima. Some aspects of this bifurcation behavior have been previously pointed out in  \cite{schwartz} and \cite{zhang-du}. Analyzing the nature of the three critical configurations for $s$ in the interval $(13.5204990011...,\, 21.1471229401...)$, Mountain Pass type theorem suggest that there should be another critical configuration with Morse index one. Indeed, we find the critical configuration of a double-tetrahedron and prove that it has Morse index one. This configuration was previously discovered in \cite{zhang-du} for various values of $s$ in this interval. We also prove that this configuration persists also for $s>21.1471229401...$.

\section{Embedded Newton algorithm}

We want to rewrite the (covariant) Newton equation \eqref{Newton-equation-forms} on the manifold $S$ in terms of the Riemannian geometry of an ambient space $M$.
Assume that there exists an ambient smooth Riemannian manifold $(M,{\bf g})$ and a smooth map ${\bf F}:M\rightarrow \R^k$ such that the manifold $S$ can be written as $S={\bf F}^{-1}(c)$, where $c\in \R^k$ is a regular value of the map ${\bf F}$. Furthermore, suppose that the induced metric on $S$ by the ambient metric ${\bf g}$ is the metric ${\bf g}_{_S}$, i.e. ${\bf g}_{_S}={\bf g}_{_{|S}}$. Let $G:M\rightarrow \R$ be a smooth prolongation of the cost function $\wt{G}$.

We fix an adapted frame $\{{\bf b}_1,...,{\bf b}_{s},\nabla F_1,...,\nabla F_k\}$ on the regular foliation induced by ${\bf F}$. More precisely, for $i=\overline{1,s}$ we have that the vector field ${\bf b}_i$ restricted to the submanifold $S$ is a tangent vector field to this submanifold, i.e. ${\bf b}_{i_{|S}}\in \mathcal{X}(S)$ and $\nabla F_{1_{|S}},...,\nabla F_{k_{|S}}$ generate the normal subspace to the submanifold $S$. 

Starting with the above adapted frame we consider a co-base $\{\Theta_1,...\Theta_s,dF_1,...,dF_k\}$ for the module of 1-forms $\Lambda^1(M)$, where for $i,j=\overline{1,s}$ we have $\Theta_i({\bf b}_j)=\delta_{ij}$ and for $\alpha=\overline{1,k}$ we have $\Theta_i(\nabla F_{\alpha})=0$.

On the manifold $S$ the equation \eqref{Newton-equation-forms} is equivalent with 
\begin{equation}\label{Newton-equation-forms-1}
 \text{Hess}\, {\wt{G}}(x_n) (\wt{{\bf v}}_{x_n}, \wt{{\bf w}}_{x_n})=-d\wt{G}(x_n)\cdot \wt{{\bf w}}_{x_n},\,\,\,\forall \wt{{\bf w}}_{x_n}\in T_{x_n}S.
\end{equation}
Regarding the tangent vectors $\wt{{\bf v}}_{x_n}, \wt{{\bf w}}_{x_n}\in T_{x_n}S$ as vectors ${\bf v}_{x_n}, {\bf w}_{x_n}\in T_{x_n}S\subset  T_{x_n}M$ in the ambient space and 
using the formula for the Hessian of a constrained function,  see  Theorem 2.1, formula 2.5 from \cite{Birtea-Comanescu-Hessian} and the Appendix {\bf 1.}, the above equation can be rewritten in the equivalent form:
\begin{equation}\label{Newton-ambient}
 \left(\text{Hess}\, G(x_n)-\sum_{\alpha=1}^k\sigma_{\alpha}(x_n)\text{Hess}\, F_{\alpha}(x_n)\right)({\bf v}_{x_n},{\bf w}_{x_n})=-dG(x_n)\cdot {\bf w}_{x_n},\,\,\forall {\bf w}_{x_n}\in T_{x_n}S\subset  T_{x_n}M,
\end{equation}
where $\sigma_{\alpha}:M\rightarrow \R$ are the Lagrange multiplier functions, see Appendix {\bf 2.} eq. \eqref{sigma}.

In the adapted co-frame, we have 
$$dG(x_n)=\sum_{i=1}^s g_i(x_n)\Theta_i(x_n)+\sum_{\alpha=1}^k f_{\alpha}(x_n)dF_{\alpha}(x_n),$$
and using the orthogonality between the vector fields $\nabla F_{\alpha}$ and ${\bf b}_j$, we obtain the formula for the coordinate functions 
\begin{equation*}\label{coefficients-g}
 g_i(x_n)=dG(x_n)\cdot {\bf b}_i(x_n).
\end{equation*}

Taking a tangent vector $\wt{{\bf w}}_{x_n}\in T_{x_n}S$ to the submanifold $S$  written as a vector in the ambient space $M$, we have $ {\bf w}_{x_n}=\sum_{j=1}^s w^j_{x_n}{\bf b}_j(x_n).$
Consequently, using again the orthogonality between the vector fields $\nabla F_{\alpha}$ and ${\bf b}_j$, for the right hand side of \eqref{Newton-ambient}, we have
\begin{align*}
dG(x_n)\cdot {\bf w}_{x_n}= & \sum_{i,j=1}^s g_i(x_n) w^j_{x_n}\Theta_i({\bf b}_j)(x_n)+\sum_{\alpha=1}^k\sum_{j=1}^s f_{\alpha}(x_n) w^j_{x_n} dF_{\alpha}({\bf b}_j)(x_n) \\
= & \sum_{i,j=1}^s g_i(x_n) w^j_{x_n}\delta_{ij}+\sum_{\alpha=1}^k\sum_{j=1}^s f_{\alpha}(x_n) w^j_{x_n}{\bf g}(\nabla F_{\alpha}(x_n),{\bf b}_j(x_n)) \\
= & \sum_{i=1}^s g_i(x_n) w^j_{x_n}.
\end{align*}
In the adapted co-frame, we have 
$$\text{Hess}\, G-\sum_{\alpha=1}^k\sigma_{\alpha}\text{Hess}\, F_{\alpha}=\sum_{i,j=1}^s h_{ij}\Theta_i\otimes\Theta_j+\sum_{i=1}^s\sum_{\alpha=1}^k \left(h_{i\alpha}\Theta_i\otimes dF_{\alpha}+h_{\alpha i}dF_{\alpha}\otimes\Theta_i\right)+\sum_{\alpha,\beta=1}^k h_{\alpha\beta}dF_{\alpha}\otimes dF_{\beta}.$$
Consequently, 
$$ \left(\text{Hess}\, G(x_n)-\sum_{\alpha=1}^k\sigma_{\alpha}(x_n)\text{Hess}\, F_{\alpha}(x_n)\right)({\bf v}_{x_n},{\bf w}_{x_n})=\sum_{i,j=1}^s h_{ij}(x_n)v^i_{x_n}w^j_{x_n},$$
where  ${\bf v}_{x_n}=\sum_{i=1}^s v^i_{x_n}{\bf b}_i(x_n)$ and $$h_{ij}(x_n)= \left(\text{Hess}\, G(x_n)-\sum_{\alpha=1}^k\sigma_{\alpha}(x_n)\text{Hess}\, F_{\alpha}(x_n)\right)({\bf b}_i({x_n}),{\bf b}_j({x_n})).$$
Erasing  the vector ${\bf w}_{x_n}$, equation \eqref{Newton-ambient} is equivalent with the system of $s$ equations with $s$ unknown variables $(v^1_{x_n},...,v^s_{x_n})$:
\begin{equation}
\begin{cases}
\begin{array}{ll}
\sum_{i=1}^s h_{i1}(x_n)v^i_{x_n}=-g_1(x_n) \\
... \\
\sum_{i=1}^s h_{is}(x_n)v^i_{x_n}=-g_s(x_n)
\end{array}%
\end{cases}
\end{equation}

The considerations above are meant to construct a numerical algorithm in order to solve the following problem.
\medskip

{\bf Problem:} Find critical points for the smooth function $\wt{G}:S\rightarrow \R$, where  $S={\bf F}^{-1}(c)$ is the preimage of a regular value for the smooth map ${\bf F}:M\rightarrow \R^k$.
\medskip

The following is the Newton algorithm on constraint manifolds written in the ambient coordinates (usually Euclidean coordinates) on the manifold $M$ (usually an Euclidean space).
\newpage

{\bf Embedded Newton algorithm}:
\begin{framed}
\begin{itemize}
\item [1.] Consider a smooth prolongation $G:M\rightarrow \R$ of the cost function  $\wt{G}:S\rightarrow \R$.

\item [2.] Construct an adapted frame $\{{\bf b}_1,...,{\bf b}_{s},\nabla F_1,...,\nabla F_k\}$, where the vector fields ${\bf b}_i\in TM$ are tangent to the submanifold $S$.

\item [3.] Compute the coordinate functions 
\begin{equation}\label{coefficients-g}
 g_i=dG\cdot {\bf b}_i,\,\,i\in\overline{1,s}.
\end{equation}

\item [4.] Compute the Lagrange multiplier functions, see Appendix eq. \eqref{sigma}, $\sigma_{\alpha}:M\rightarrow \R$, for $\alpha=\overline{1,k}$.

\item [5.] Compute the components of the Hessian matrix $\text{Hess}\,\wt{G}$ of the cost function $\wt{G}$
\begin{equation}\label{HessR}
 h_{ij}= \left(\text{Hess}\, G-\sum_{\alpha=1}^k\sigma_{\alpha}\text{Hess}\, F_{\alpha}\right)({\bf b}_i,{\bf b}_j),\,\,i,j\in \overline{1,s}.
\end{equation}

\item [6.] Choose a retraction $\mathcal{R}:T_{x}M\rightarrow M$ such that for any ${\bf v}\in T_{x}S\subset T_{x}M$ we have $\mathcal{R}_{x}({\bf v})\in S$.

\item [7.] Input ${\bf x}_0\in S$ and $n=0$.

\item [8.] {\bf repeat}

$\bullet$ Solve the linear system with the unknowns  $(v^1_{x_n},...,v^s_{x_n})$,
\begin{equation}\label{newton-equation}
\begin{cases}
\sum_{i=1}^s h_{i1}(x_n)v^i_{x_n}=-g_1(x_n) \\
... \\
\sum_{i=1}^s h_{is}(x_n)v^i_{x_n}=-g_s(x_n).
\end{cases}
\end{equation}

$\bullet$ Construct the line search vector $$ {\bf v}_{x_n}=\sum_{j=1}^s v^j_{x_n}{\bf b}_j(x_n).$$

$\bullet$ Set $x_{n+1}=\mathcal{R}_{x_n}({\bf v}_{x_n})$.

{\bf until} $x_{n+1}$ sufficiently minimizes $\wt{G}$.

\end{itemize}
\end{framed}

\section{5-electron Thomson problem}\label{thomson}

We will apply the above numerical algorithm to the problem of finding critical configurations for the 5-electron Thomson problem.
The 5-electron Thomson problem is the following: consider 5 electrons constrained on a unit sphere interacting through Coulomb force. Which are the configurations that renders the Coulomb potential its minimum value? 
The mathematical 5-electron Thomson problem is:
\begin{equation}
\text{argmin}_{||{\bf p}_i||=1}\sum_{1\leq i<j\leq 5}\frac{1}{||{\bf p}_i-{\bf p}_j||}, 
\end{equation}
where ${\bf p}_i$ is the position vector of the electron ${\bf P}_i$ on the unit sphere. 

Without minimizing the generality of the problem we can suppose that the electron ${\bf P}_5$ is fixed in the North pole, i.e. ${\bf p}_5:=(0,0,1)$.
The phase space of the problem is the manifold
$$S:=S^2\times S^2\times S^2\times S^2\backslash \{{\bf p}:=({\bf p}_1,{\bf p}_2,{\bf p}_3,{\bf p}_4)\,|\,\exists (i,j), 1\leq i<j\leq 5,\,\text{s.t.}\,{\bf p}_i={\bf p}_j\}.$$
The cost function $\wt{G}:S\rightarrow \R$ is a Coulomb potential and it is given by 
$$\wt{G}({\bf p}):=\sum_{1\leq i<j\leq 5}\frac{1}{||{\bf p}_i-{\bf p}_j||}.$$

The aim is to find all critical configurations of the above Coulomb potential. First we will proceed by constructing Embedded Newton algorithm for the constraint configuration space $S$. This will give us numerically three types of critical configurations. We also make an analytical study of this configurations. 

\subsection{Numerical study: Embedded Newton algorithm} We embed this optimization problem into $\R^{12}$ using the constraint functions ${\bf F}:=(F_1,F_2,F_3,F_4):M\rightarrow \R^4$, where the components are $F_i ({\bf p}):=\frac{1}{2}||{\bf p}_i||^2$, $i=\overline{1,4}$ and the ambient space $M$ is the open set
$\R^{12}\backslash \{{\bf p}\,|\,\exists (i,j), 1\leq i<j\leq 5,\,\text{s.t.}\,{\bf p}_i={\bf p}_j\}$. The extended cost function $G:M\rightarrow \R$ is given by $$G({\bf p}):=\sum_{1\leq i<j\leq 5}\frac{1}{||{\bf p}_i-{\bf p}_j||}.$$

On the unit sphere $S^2\subset \R^3$ we can construct a local frame using the North pole stereographic projection: ${\bf e}_1(x,y,z):=
(1-z-x^2,-xy,x(1-z))$, ${\bf e}_2(x,y,z):=(-xy,1-z-y^2,y(1-z))$, see Appendix {\bf 3.}.

A local frame on the constraint manifold $S$ is given by:
$${\bf b}_1({\bf p}):=({\bf e}_1({\bf p}_1),{\bf 0},{\bf 0},{\bf 0}), \,\,{\bf b}_2({\bf p}):=({\bf e}_2({\bf p}_1),{\bf 0},{\bf 0},{\bf 0}),$$
$${\bf b}_3({\bf p}):=({\bf 0},{\bf e}_1({\bf p}_2),{\bf 0},{\bf 0}), \,\,{\bf b}_4({\bf p}):=({\bf 0},{\bf e}_2({\bf p}_2),{\bf 0},{\bf 0}),$$
$${\bf b}_5({\bf p}):=({\bf 0}, {\bf 0},{\bf e}_1({\bf p}_3),{\bf 0}), \,\,{\bf b}_6({\bf p}):=({\bf 0},{\bf 0},{\bf e}_2({\bf p}_3),{\bf 0}),$$
$${\bf b}_7({\bf p}):=({\bf 0},{\bf 0}, {\bf 0},{\bf e}_1({\bf p}_4)), \,\,{\bf b}_8({\bf p}):=({\bf 0},{\bf 0},{\bf 0},{\bf e}_2({\bf p}_4)).$$

The eight coordinate functions are given by:
$$g_1({\bf p})=<\frac{\partial G}{\partial {\bf p}_1}({\bf p}),{\bf e}_1({\bf p}_1)>,\,\,g_2({\bf p})=<\frac{\partial G}{\partial {\bf p}_1}({\bf p}),{\bf e}_2({\bf p}_1)>,$$
$$g_3({\bf p})=<\frac{\partial G}{\partial {\bf p}_2}({\bf p}),{\bf e}_1({\bf p}_2)>,\,\,g_4({\bf p})=<\frac{\partial G}{\partial {\bf p}_2}({\bf p}),{\bf e}_2({\bf p}_2)>,$$
$$g_5({\bf p})=<\frac{\partial G}{\partial {\bf p}_3}({\bf p}),{\bf e}_1({\bf p}_3)>,\,\,g_6({\bf p})=<\frac{\partial G}{\partial {\bf p}_3}({\bf p}),{\bf e}_2({\bf p}_3)>,$$
$$g_7({\bf p})=<\frac{\partial G}{\partial {\bf p}_4}({\bf p}),{\bf e}_1({\bf p}_4)>,\,\,g_8({\bf p})=<\frac{\partial G}{\partial {\bf p}_4}({\bf p}),{\bf e}_2({\bf p}_4)>.$$

By a straightforward computation we obtain the Lagrange multipliers functions:
$$\sigma_1({\bf p}):=<\frac{\partial G}{\partial {\bf p}_1}({\bf p}),{\bf p}_1)>, \,\,\sigma_2({\bf p}):=<\frac{\partial G}{\partial {\bf p}_2}({\bf p}),{\bf p}_2)>,$$ 
$$\sigma_3({\bf p}):=<\frac{\partial G}{\partial {\bf p}_3}({\bf p}),{\bf p}_3)>, \,\,\sigma_4({\bf p}):=<\frac{\partial G}{\partial {\bf p}_4}({\bf p}),{\bf p}_4)>.$$ 

We have all the necessary elements  to compute the Hessian matrix of the constraint function $\wt{G}$ applying formula \eqref{HessR}. The expressions are long and irrelevant for what follows and we choose not to write them down.

We need to construct a retract on $M$. A natural choice is $\mathcal{\wt{R}}_{\bf p}:T_{\bf p}M\rightarrow M$ given by 
$$\mathcal{\wt{R}}_{\bf p}({\bf v}_{\bf p}):=(\mathcal{R}_{{\bf p}_1}({\bf v}_{{\bf p}_1}),...,\mathcal{R}_{{\bf p}_4}({\bf v}_{{\bf p}_4})),$$
where ${\bf v}_{{\bf p}_i}\in T_{{\bf p}_i}S^2$, ${\bf v}_{\bf p}=({\bf v}_{{\bf p}_1},...,{\bf v}_{{\bf p}_4})$, and $\mathcal{R}_{{\bf p}_i}({\bf v}_{{\bf p}_i})=\frac{{\bf p}_i+{\bf v}_{{\bf p}_i}}{|| {\bf p}_i+{\bf v}_{{\bf p}_i}||}$ is the usual retraction on the sphere.
For a detailed discussion of retractions and their use see \cite{absil-book}.

After running many numerical experiments using the Newton algorithm described above we find three types of configurations that are critical points for the cost function $\wt{G}$.

{\bf 1. Bi-pyramidal configuration.}  In this configuration two points are opposed one to another and the other three lie on an equilateral triangle in a plane perpendicular on the diameter formed by the first two points. For our problem we find two types of bi-pyramidal configurations, one in which ${\bf P}_5$ is one of the vertex of the equilateral triangle and other one is when ${\bf P}_5$ is not a vertex of the equilateral triangle but one of diametrically opposed points. The value of the cost function $\wt{G}$ on both types of bi-pyramidal configurations is the same and it is equal with the known value $\frac{1}{2}+3\sqrt{2}+\sqrt{3}=6.47461494...$.

{\bf 2. Square right pyramid.} This configuration has a right square at the base located at a distance of $1+0.2432010309...$ from the apex. The value of the cost function $\wt{G}$ on this configuration is equal with $6.483660519....$

{\bf 3. Regular pentagon.} In this configurations the five electrons form a regular pentagon which is inscribed in a great circle that goes to North pole. The value of the cost function $\wt{G}$ on this configuration is equal with $6.881909602....$
\medskip

The bi-pyramidal configurations are the well known local minima and indeed, we find that the Hessian matrix of the cost function $\wt{G}$ in such a configuration has seven strictly positive eigenvalues and one eigenvalue is equal with zero. In \cite{schwartz} it has been given a rigorous computer-assisted proof that this configurations are the only global minima. See also \cite{huo-shao} for a computer-assisted proof for the solution of a similar problem.

For the square right pyramid configuration we find that the Hessian matrix of the cost function $\wt{G}$ in such a configuration has six strictly positive eigenvalues, one strictly negative eigenvalue, and one eigenvalue is equal with zero. This configuration has been extensively studied in \cite{bondarenko}, \cite{melnyk}, \cite{zhang-du}.

For the regular pentagon configuration we find that the Hessian matrix of the cost function $\wt{G}$ in such a configuration has five strictly positive eigenvalues, two strictly negative eigenvalues, and one eigenvalue is equal with zero. 

We will show in the next section that the zero eigenvalue is in fact a degeneracy that can be eliminated using Morse-Bott Theorem and that all three configurations are nondegenerate in the sense of Morse-Bott theory.

\subsection{Analytical study}
In this section we will show that for all three types of configurations, each belongs to a differential curve of critical configurations. Next we will compute the Hessian matrix of the cost function $\wt{G}$ in this critical configurations and study their nature. For all this three types of critical configurations, the Hessian matrix has one eigenvalue  equal to zero. We will prove that this is due to the fact that these configurations belong to one dimensional submanifolds of critical configurations. We will apply the Morse-Bott theory to study them.

To prove that a configuration is critical we have to verify the equation $d\wt{G}({\bf p})=0$. In order to do this we have to introduce local coordinates on the manifold $S$, see \cite{muller}. An alternative method is to embed $S$ in the ambient space $M$ and solve an equivalent equation written in Cartesian coordinates. In \cite{birtea-comanescu} and \cite{Birtea-Comanescu-Hessian} we have introduced and studied the vector field on the ambient space $M$
\begin{equation}\label{v0-grad}
\partial G({\bf p}) =\nabla G({\bf p})-\sum_{i=1}^4\sigma_i({\bf p})\nabla F_i({\bf p}),
\end{equation}
which has the property that when restricted to the submanifold $S$ it is equal with the gradient of the cost function $\wt{G}$ with respect to the induced metric, i.e. $(\partial G)_{|S}=\nabla _{g_{ind}^S}\wt{G}$. Due to this property we will call the vector field $\partial G$ the {\bf embedded gradient vector field}. To verify that {\bf a configuration is critical for $\wt{G}$ is equivalent with verifying the equation $\partial G({\bf p})=0$}, which is an equation written in Cartesian coordinates. This technique has also been used in \cite{birtea-comanescu-popa}.

For the case of the constraint functions $F_1,...,F_4$, a direct computation shows that the equation $\partial G({\bf p})=0$ is equivalent with the following system:
\begin{equation}
\partial G({\bf p})={\bf 0}\,\,\Leftrightarrow \,\,\frac{\partial G }{\partial {\bf p_i}}-<\frac{\partial G }{\partial {\bf p_i}},{\bf p}_i>{\bf p}_i={\bf 0},\,\,\,i=\overline{1,4}.
\end{equation} 
For the particular case of the Coulomb potential we have the following system of twelve scalar equations:
\begin{equation}
\partial G({\bf p})={\bf 0}\,\,\Leftrightarrow \,\,\sum_{j=1,\,j\neq i}^5\frac{{\bf p}_j-<{\bf p}_j,{\bf p}_i>{\bf p}_i}{||{\bf p}_i-{\bf p}_j||^3}=0,\,\,i=\overline{1,4}.
\end{equation}

Next we will give analytical expressions for the three types of critical configurations previously discovered numerically. 
\medskip

{\bf 1.  Bi-pyramidal configuration}. 

Let $c:[-1,1] \rightarrow M$ be the differential curve $c(\lambda):=({\bf P}_1(\lambda),{\bf P}_2(\lambda),{\bf P}_3(\lambda),{\bf P}_4(\lambda),{\bf P}_5)$, where 
\begin{equation}\label{bi-pyramidal-critical-points}
\begin{cases}
{\bf P}_1(\lambda)=(\frac{\sqrt{3}}{2}\lambda,-\frac{\sqrt{3}}{2}\sqrt{1-\lambda^2},-\frac{1}{2}) \\
{\bf P}_2(\lambda)=(-\frac{\sqrt{3}}{2}\lambda,\frac{\sqrt{3}}{2}\sqrt{1-\lambda^2},-\frac{1}{2}) \\
{\bf P}_3(\lambda)=(\sqrt{1-\lambda^2},\lambda,0) \\
{\bf P}_4(\lambda)=(-\sqrt{1-\lambda^2},-\lambda,0) \\
\end{cases}.
\end{equation}
By direct computation we obtain that $\partial G(c(\lambda))=0$, for all $\lambda\in [-1,1]$. Consequently, $c$ is a curve of critical points for the Coulomb potential $\wt{G}$. Every point of this curve generates a bi-pyramidal configuration with ${\bf P}_5$ being one of the vertex of the equilateral triangle.

Using formula \eqref{HessR} we obtain the eight eigenvalues of the Hessian matrix $\text{Hess}\,\wt{G}(c(\lambda))$ of the Coulomb potential as follows: 
\begin{align*}
\lambda_1 & =0 \\
\lambda_2 & = {\frac {9}{32}}\,\sqrt {2}+\frac{1}{8}+{\frac {15}{32}}\,\sqrt {3}+\frac{1}{32}\,
\sqrt {1717+72\,\sqrt {2}-270\,\sqrt {2}\sqrt {3}-120\,\sqrt {3}} =2.297...\\
\lambda_3 & = {\frac {9}{32}}\,\sqrt {2}+\frac{1}{8}+{\frac {15}{32}}\,\sqrt {3}-\frac{1}{32}\,
\sqrt {1717+72\,\sqrt {2}-270\,\sqrt {2}\sqrt {3}-120\,\sqrt {3}} =0.371...\\
\lambda_4 & = \frac{1}{8}+{\frac {9}{32}}\,\sqrt {2}+{\frac {5}{32}}\,\sqrt {3}+\frac{1}{32}\,\sqrt 
{541+72\,\sqrt {2}-40\,\sqrt {3}-90\,\sqrt {2}\sqrt {3}}=1.380... \\
\lambda_5 & = \frac{1}{8}+{\frac {9}{32}}\,\sqrt {2}+{\frac {5}{32}}\,\sqrt {3}-\frac{1}{32}\,\sqrt 
{541+72\,\sqrt {2}-40\,\sqrt {3}-90\,\sqrt {2}\sqrt {3}}=0.206... \\
\lambda_6 & =\frac{1}{4}\,\sqrt {3}+{\frac {9}{8}}\,\sqrt {2}+\frac{1}{8}\,\sqrt {93+18\,\sqrt {2}
\sqrt {3}}=3.487...\\
\lambda_7 & =\frac{1}{4}\,\sqrt {3}+{\frac {9}{8}}\,\sqrt {2}-\frac{1}{8}\,\sqrt {93+18\,\sqrt {2}
\sqrt {3}}=0.560...\\
\lambda_8 & =\frac{9\sqrt{2}}{4}=3.181...
\end{align*}

We prove that the eigenvalue $\lambda_1=0$ is a manifestation of the Morse-Bott Theory, see \cite{bott} and \cite{nicolaescu}. The connected and compact set $\mathcal{C}:=c([-1,1])\subset S$ of critical points of Coulomb potential $\wt{G}$ is a 1-dimensional submanifold of $S$. We prove that $T_{c(\lambda)}\mathcal{C}=\text{Ker}[\text{Hess}\,\wt{G}(c(\lambda))]$, for all $\lambda\in [-1,1]$. For this we write the tangent vector $c'(\lambda)\in \R^{12}$ as a vector in $T_{c(\lambda)}S\simeq \R^8$, i.e. solving the equation 
$c'(\lambda)=\sum_{i=1}^8w_i(\lambda){\bf b}_i(c(\lambda))$
for unknowns $w_i$. The solution is given by the tangent vector
$${\bf w}(\lambda)=(\frac{\sqrt{3}}{3}, \frac{\sqrt{3}\lambda}{3\sqrt{1-\lambda^2}}, -\frac{\sqrt{3}}{3}, -\frac{\sqrt{3}\lambda}{3\sqrt{1-\lambda^2}}, -\frac{\lambda}{\sqrt{1-\lambda^2}}, 1,\frac{\lambda}{\sqrt{1-\lambda^2}}, -1)\in T_{c(\lambda)}S.$$
A direct computation shows that 
$\text{Hess}\,\wt{G}(c(\lambda))\cdot {\bf w}(\lambda)={\bf 0},\,\,\forall\,\lambda\in (-1,1)$, which is the condition of Morse-Bott Theorem, that is the eigenspace corresponding to the eigenvalue $\lambda_1=0$ is equal with the tangent space of the submanifold of critical points $\mathcal{C}$ that generate the bi-pyramidal configuration. Because all other eigenvalues are strictly positive we have that the bi-pyramidal configuration are non-degenerate in the sense of Morse-Bott local minima.

Any relabeling in \eqref{bi-pyramidal-critical-points} will give connected, compact 1-dimensional submanifolds of critical points that will generates the same bi-pyramidal configuration  with ${\bf P}_5$ is one of the vertex of the equilateral triangle. Their study is analogous, all of them being non-degenerate in the sense of Morse-Bott local minima. Also the bi-pyramidal configurations where ${\bf P}_5$ is not one of the vertex of the equilateral triangle come from 1-dimensional submanifolds of critical points and are non-degenerate in the sense of Morse-Bott local minima. 
\medskip

{\bf 2. Square right pyramid.}
As the numerical study shows we have a critical configuration of square right pyramid where ${\bf P}_5$ is the apex and the base is at the distance $1+0.2432010309...$. We prove analytically this is indeed true. 

Let $c_{\alpha}:[0,2{\pi}] \rightarrow M$ be the differential curve $c_{\alpha}(\lambda):=({\bf P}^{\alpha}_1(\lambda),{\bf P}^{\alpha}_2(\lambda),{\bf P}^{\alpha}_3(\lambda),{\bf P}^{\alpha}_4(\lambda),{\bf P}_5)$, where
\begin{equation}\label{rectangular-pyramid-critical-points}
\begin{cases}
{\bf P}^{\alpha}_1(\lambda)=(\sqrt{1-\alpha^2}\cos(\lambda),\sqrt{1-\alpha^2}\sin(\lambda),\alpha) \\
{\bf P}^{\alpha}_2(\lambda)=(\sqrt{1-\alpha^2}\cos(\lambda+\frac{\pi}{2}),\sqrt{1-\alpha^2}\sin(\lambda+\frac{\pi}{2}),\alpha) \\
{\bf P}^{\alpha}_3(\lambda)=(\sqrt{1-\alpha^2}\cos(\lambda+\pi),\sqrt{1-\alpha^2}\sin(\lambda+\pi),\alpha) \\
{\bf P}^{\alpha}_4(\lambda)=(\sqrt{1-\alpha^2}\cos(\lambda+\frac{3\pi}{2}),\sqrt{1-\alpha^2}\sin(\lambda+\frac{3\pi}{2}),\alpha)
\end{cases},
\end{equation}
and $|\alpha|\in [0,1)$ is the distance from the center of the sphere to the center of the square base of the pyramid. 
By direct computation we obtain that 
\begin{align*}
\partial G(c_{\alpha}(\lambda))= -\frac{\alpha}{2\sqrt{1-\alpha^2}}Q(\alpha) & \Bigg(\cos(\lambda),\sin(\lambda),-\frac{\sqrt{1-\alpha^2}}{\alpha},-\sin(\lambda),\cos(\lambda),-\frac{\sqrt{1-\alpha^2}}{\alpha}, \\
& -\cos(\lambda),-\sin(\lambda),-\frac{\sqrt{1-\alpha^2}}{\alpha},\sin(\lambda),-\cos(\lambda),-\frac{\sqrt{1-\alpha^2}}{\alpha}\Bigg),
\end{align*}
for all $\lambda\in [0,2{\pi}]$, $\alpha\in (-1,1)$, and 
$$Q(\alpha)=\left( 2-2\,\alpha \right) ^{-\frac{1}{2}}+2\, \left( 2-2\,{\alpha}^{2}
 \right) ^{-\frac{1}{2}}\alpha+\alpha\, \left( 4-4\,{\alpha}^{2} \right) ^{
-\frac{1}{2}}+ \left( 2-2\,\alpha \right) ^{-\frac{1}{2}}\alpha.$$
Consequently, $\partial G(c_{\alpha}(\lambda))=0$ iff $Q(\alpha)=0$. The only solution in the interval $(-1,1)$ is $\alpha^*=-0.24320103..$
Using formula \eqref{HessR} we obtain the eight eigenvalues of the Hessian matrix $\text{Hess}\,\wt{G}(c_{\alpha^*}(\lambda))$ of the Coulomb potential as follows: 
$$\lambda_1=0;\,\,\lambda_2=-0.205..;\,\lambda_3=0.565..;\,\lambda_4=0.565..;\lambda_5=2.232..;\,\lambda_6=2.232..;\,\lambda_7=2.260..;\,\lambda_8=3.592..$$
As before, we show that we have a Morse-Bott non-degeneracy. A straightforward computation shows that 
$\left(\text{Hess}\, G(c_{\alpha^*}(\lambda))-\sum_{i=1}^4\sigma_{i}(c_{\alpha^*}(\lambda))\text{Hess}\, F_{i}(c_{\alpha^*}(\lambda))+\right)\cdot c_{\alpha^*}'(\lambda)=0$, which implies that the tangent vector $c_{\alpha^*}'(\lambda)$ to the critical submanifold $c_{\alpha^*}([0,2{\pi}])$ is also an eigenvector corresponding to the eigenvalue $\lambda_1=0$. Thus the square right pyramid configurations are non-degenerate in the sense of Morse-Bott saddle points for the Coulomb potential $\wt{G}$.

 Any relabeling in \eqref{rectangular-pyramid-critical-points} will give connected, compact 1-dimensional submanifolds of critical points that will generate the same square right pyramid configuration  with ${\bf P}_5$ the apex. Their study is analogous, all of them being non-degenerate in the sense of Morse-Bott saddle points. We also find square right pyramid critical configurations where ${\bf P}_5$ is one of the vertex for the base square of the pyramid. These configurations are also nondegenerate Morse-Bott saddle points with same value $6.483660519....$ of the Coulomb potential.
\medskip

{\bf 3. Regular pentagon.} The numerical study shows that we obtain a configuration of critical points in the shape of a regular pentagon sitting on the big circles that go through ${\bf P}_5$ that is fixed in the North pole. This configuration seems to be new, at least the authors could not find it in the literature.

Let $c:[0,2\pi] \rightarrow M$ be the differential curve $c(\lambda):=({\bf P}_1(\lambda),{\bf P}_2(\lambda),{\bf P}_3(\lambda),{\bf P}_4(\lambda),{\bf P}_5)$, where 
\begin{equation}\label{pentagon-critical-points}
\begin{cases}
{\bf P}_1(\lambda)=(-\sin(\lambda)\cos(\frac{\pi}{10}),-\cos(\lambda)\cos(\frac{\pi}{10}),\sin(\frac{\pi}{10})) \\
{\bf P}_2(\lambda)=(-\sin(\lambda)\cos(\frac{3\pi}{10}),-\cos(\lambda)\cos(\frac{3\pi}{10}),-\sin(\frac{3\pi}{10})) \\
{\bf P}_3(\lambda)=(\sin(\lambda)\cos(\frac{3\pi}{10}),\cos(\lambda)\cos(\frac{3\pi}{10}),-\sin(\frac{3\pi}{10}))  \\
{\bf P}_4(\lambda)=(\sin(\lambda)\cos(\frac{\pi}{10}),\cos(\lambda)\cos(\frac{\pi}{10}),\sin(\frac{\pi}{10})) 
\end{cases}.
\end{equation}

By direct computation we obtain that $\partial G(c(\lambda))=0$, for all $\lambda\in [0,2\pi]$. Consequently, $c$ is a curve of critical points for the Coulomb potential $\wt{G}$. Every point of this curve generates a pentagonal configuration.

Using formula \eqref{HessR} we obtain the eight eigenvalues of the Hessian matrix $\text{Hess}\,\wt{G}(c(\lambda))$ of the Coulomb potential as follows: 
$$\lambda_1=0;\lambda_2=-2.628..;\lambda_3=-0.453..;\lambda_4=0.490..;\lambda_5=1.084..;\lambda_6=1.992..;\lambda_7=4.932..;\lambda_8=11.156..$$
A straightforward computation shows that 
$\left(\text{Hess}\, G-\sum_{i=1}^4\sigma_{i}\text{Hess}\, F_{i}\right)\cdot c'(\lambda)=0$ which implies that the tangent vector $c'(\lambda)$ to the critical submanifold $c([0,2\pi])$ is also an eigenvector corresponding to the eigenvalue $\lambda_1=0$. Thus the pentagonal configurations are non-degenerate in the sense of Morse-Bott saddle points for the Coulomb potential $\wt{G}$.

\section{Riesz s-energy of 5-points on a sphere}\label{Riesz}
A generalization of 5-electron Thomson problem is given by the minimization problem of Riesz s-energy of 5-points on a sphere:
\begin{equation}
\text{argmin}_{||{\bf p}_i||=1}\sum_{1\leq i<j\leq 5}\frac{1}{||{\bf p}_i-{\bf p}_j||^s}, 
\end{equation} 
where we use the notations of Section \ref{thomson} and $s> 0$ represents for example the soft or strong repulsion in VSEPR (Valence Shell Electron Pair Repulsion) model, see \cite{gillespie}, \cite{gillespie-hargitai}, and \cite{hargitai-chamberland}. A more general problem of finding optimal configurations using general potentials on $d$-dimensional spheres is studied in \cite{ball}, \cite{cohn}, and \cite{cohn-kumar}.

In this section we study the critical points of the Riesz s-energy of 5 points on the sphere $\wt{G}_s:S\rightarrow \R$ given by 
$$\wt{G}_s({\bf p}):=\sum_{1\leq i<j\leq 5}\frac{1}{||{\bf p}_i-{\bf p}_j||^s},$$
 where the constraint manifold is
$S:=S^2\times S^2\times S^2\times S^2\backslash \{{\bf p}:=({\bf p}_1,{\bf p}_2,{\bf p}_3,{\bf p}_4)\,|\,\exists (i,j), 1\leq i<j\leq 5,\,\text{s.t.}\,{\bf p}_i={\bf p}_j\}.$

Using the same embedding mechanism as in Section \ref{thomson} we obtain that the critical points of the Riesz s-energy $\wt{G}_s$ are the solutions of the system of twelve scalar equations:
 \begin{equation}\label{v0-riesz}
\partial G_s({\bf p})={\bf 0}\,\,\Leftrightarrow \,\,\sum_{j=1,\,j\neq i}^5\frac{{\bf p}_j-<{\bf p}_j,{\bf p}_i>{\bf p}_i}{||{\bf p}_i-{\bf p}_j||^{s+2}}=0,\,\,i=\overline{1,4},
\end{equation}
where $G_s$ is the natural extension of $\wt{G}_s$ on $M=\R^{12}\backslash \{{\bf p}\,|\,\exists (i,j), 1\leq i<j\leq 5,\,\text{s.t.}\,{\bf p}_i={\bf p}_j\}$.
\medskip

{\bf 1. Bi-pyramid configuration.} 
By direct computation we obtain that the bi-pyramid configuration \eqref{bi-pyramidal-critical-points} verifies equation \eqref{v0-riesz} and consequently, it remains a critical configuration of $\wt{G}_s$ for any $s>0$ with $$\wt{G}_s(\text{bi-pyramid})=\frac{3}{\sqrt{3}^s}+\frac{6}{\sqrt{2}^s}+\frac{1}{2^s}.$$ 
 Computing the Hessian of $\wt{G}_s$ in this bi-pyramid configuration we obtain that for $s<21.1471229401...$ we have non-degenerate local minima in the sense of Morse-Bott. This has been previously noticed and discussed in \cite{schwartz} and \cite{zhang-du}. For $s>21.1471229401...$ we obtain non-degenerate saddle points in the sense of Morse-Bott with two strictly negative eigenvalues, which shows that a bifurcation occurs in this type of critical configuration.
\medskip

{\bf 2. Square right pyramid configuration.} As in the case of Thomson problem $s=1$ discussed in the previous section we obtain that the equation \eqref{v0-riesz} is equivalent for this configuration with the following equation:
\begin{equation}\label{eq-alpha}
\partial G_s(c_{\alpha}(\lambda))=0\,\Leftrightarrow \,T_s(\alpha)=0,
\end{equation}
where 
$$T_s(\alpha)=\left( 2-2\,\alpha \right) ^{-\frac{s}{2}}+2\, \left( 2-2\,{\alpha}^{2}
 \right) ^{-\frac{s}{2}}\alpha+\alpha\, \left( 4-4\,{\alpha}^{2} \right) ^{
-\frac{s}{2}}+ \left( 2-2\,\alpha \right) ^{-\frac{s}{2}}\alpha.
$$
We observe that we have $T_1(\alpha)=Q(\alpha)$, where $Q(\alpha)$ has been defined in the previous section. For every $s\geq 1$ we obtain that the equation \eqref{eq-alpha} has a solution $\alpha$ in the interval $[-0.2432010309...,0)$. When $s$ become bigger the base of the pyramid approaches the equator. For $s\in (0,1]$ the solution $\alpha$ of the equation \eqref{eq-alpha} is in the interval $(-0.25,-0.2432010309...]$ meaning that when $s$ becomes very small the base of the pyramid is at the distance close to $0.25$ from the equatorial plane.
The value of the Riesz s-energy on this critical configuration is given by 
$$\wt{G}_s(\text{square right pyramid})=\frac{2}{(\sqrt{4-4\alpha^2})^s}+\frac{4}{(\sqrt{2-2\alpha^2})^s}+\frac{4}{(\sqrt{2-2\alpha})^s},$$
 where $\alpha$ is the solution of the equation \eqref{eq-alpha}.

The study of the Hessian matrix shows that the right square pyramid configuration undergoes various bifurcation phenomena. More precisely for $s<13.5204990011...$ we have non-degenerate saddle points in the sense of Morse-Bott having one eigenvalue strictly negative. For $s>13.5204990011...$ the right square pyramid configurations are non-degenerate local minima in the sense of Morse-Bott.

For $s$ between 13.5204990011... and  15.048077392... we have that both bi-pyramidal configuration and square right pyramid configuration are local minima with 
$$\wt{G}_s(\text{bi-pyramid})<\wt{G}_s(\text{square right pyramid}).$$
For $s$ between 15.048077392... and 21.1471229401... we have that both bi-pyramidal configuration and square right pyramid configuration are local minima with 
$$\wt{G}_s(\text{bi-pyramid})>\wt{G}_s(\text{square right pyramid}),$$
which is a phenomenon discovered in \cite{schwartz}.

For $s>21.1471229401...$ the above inequality remains true.
\medskip

{\bf 3. Regular pentagon.} The regular pentagon configuration discussed in the previous section is a solution for the equation \eqref{v0-riesz} and consequently, remains critical configuration of Riesz s-energy for any $s>0$. The regular pentagon configuration remains non-degenerate saddle in the sense of Morse-Bott with two strictly negative eigenvalues for any $s>0$. 
The value of Riesz s-energy on these critical configuration has the formula
$$\wt{G}_s(\text{regular pentagon})=\frac{3}{\left(\sqrt{2-2\cos\frac{2\pi}{5}}\right)^s}+
\frac{3}{\left(\sqrt{2+2\cos\frac{\pi}{5}}\right)^s}+
\frac{2}{\left(\sqrt{2-2\sin\frac{\pi}{10}}\right)^s}
+ \frac{2}{\left(\sqrt{2+2\sin\frac{3\pi}{10}}\right)^s}.
$$

{\bf 4. Double-tetrahedron.} For $s$ in the interval $(13.5204990011...,\, 21.1471229401...)$ we have two types of local minima configurations, namely the bi-pyramid and square right pyramid configurations. Mountain Pass type theorem suggest that for $s$ in this interval one should have another critical configuration of mountain pass type, precisely critical points with Morse index at most one, see \cite{ambrosetti} and \cite{rabinowitz}. The regular pentagon configuration is a critical configuration with Morse index two and so does not fill the bill. 
Such a configuration with Morse index one was found in \cite{zhang-du} for some values of $s$ and has the shape of a double-tetrahedron.
Having this in mind, we search for critical configurations of the following type:  the segment ${\bf P}_1{\bf P}_3$ is at the distance $|\beta|$ and parallel with the equatorial plane,  the segment ${\bf P}_2{\bf P}_4$ is at the distance $|\gamma|$ and parallel with the equatorial plane, and also ${\bf P}_1{\bf P}_3\perp {\bf P}_2{\bf P}_4$.
Such a configuration is described by the differential curve $c_{\beta,\gamma}:[0,2\pi] \rightarrow M$,  $c_{\beta,\gamma}(\lambda):=({\bf P}^{\beta,\gamma}_1(\lambda),{\bf P}^{\beta,\gamma}_2(\lambda),{\bf P}^{\beta,\gamma}_3(\lambda),{\bf P}^{\beta,\gamma}_4(\lambda),{\bf P}_5)$, where
\begin{equation}\label{rectangular-pyramid-critical-points}
\begin{cases}
{\bf P}^{\beta,\gamma}_1(\lambda)=(\sqrt{1-\beta^2}\cos(\lambda),\sqrt{1-\beta^2}\sin(\lambda),\beta) \\
{\bf P}^{\beta,\gamma}_2(\lambda)=(\sqrt{1-\gamma^2}\cos(\lambda+\frac{\pi}{2}),\sqrt{1-\gamma^2}\sin(\lambda+\frac{\pi}{2}),\gamma) \\
{\bf P}^{\beta,\gamma}_3(\lambda)=(\sqrt{1-\beta^2}\cos(\lambda+\pi),\sqrt{1-\beta^2}\sin(\lambda+\pi),\beta) \\
{\bf P}^{\beta,\gamma}_4(\lambda)=(\sqrt{1-\gamma^2}\cos(\lambda+\frac{3\pi}{2}),\sqrt{1-\gamma^2}\sin(\lambda+\frac{3\pi}{2}),\gamma)
\end{cases},
\end{equation}
and $\beta,\gamma\in (-1,1)$.

The equation \eqref{v0-riesz} is equivalent for this configuration with the following system:
\begin{equation}\label{eq-beta-gamma}
\partial G_s(c_{\beta,\gamma}(\lambda))=0\,\Leftrightarrow \,
\begin{cases}
E_s(\beta,\gamma)=0 \\
E_s(\gamma,\beta)=0,
\end{cases}
\end{equation}
where 
\begin{align*}
E_s(\beta, \gamma)= & - \left( 2-2\beta \right) ^{-\frac{s}{2}}\beta- \left( 2-2\beta
 \right) ^{-\frac{s}{2}}-\beta\left( 4-4{\beta}^{2} \right) ^{-\frac{s}{2}}
+ \left( 2-2\beta \right) ^{-\frac{s}{2}}{\beta}^{2}\gamma  -2 \left( 2-2\beta\gamma \right) ^{-\frac{s}{2}}\gamma \\
& +2\left( 2-2\beta\gamma
 \right) ^{-\frac{s}{2}}{\beta}^{2}\gamma+ \left( 2-2\beta \right) ^{-\frac{s}{2}}\beta\gamma+{\beta}^{2}\gamma\, \left( 4-4{\beta}^{2} \right) 
^{-\frac{s}{2}}.
\end{align*}
An easy observation is that if the system \eqref{eq-beta-gamma} has a solution $(\beta,\gamma)$ then $(\gamma,\beta)$ is again a solution and consequently, it is sufficient to search for solutions with $\beta\geq \gamma$. 
This system has among the solutions the following pairs $(-0.5,0)$ and $(0,-0.5)$ which correspond to bi-pyramidal configurations with ${\bf P}_5$ on the equilateral triangle. If $\beta=\gamma$ we obtain the equality $E_s(\beta,\beta)=(\beta^2-1)T_s(\beta),$ where $T_s$ is the polynomial in \eqref{eq-alpha} whose solutions describes the distance from the equatorial plane of the base of the square right pyramid configuration. Thus we obtain again the square right pyramid configuration as critical configuration.

\begin{figure}[h!]
  \centering
    \includegraphics[width=0.95\textwidth]{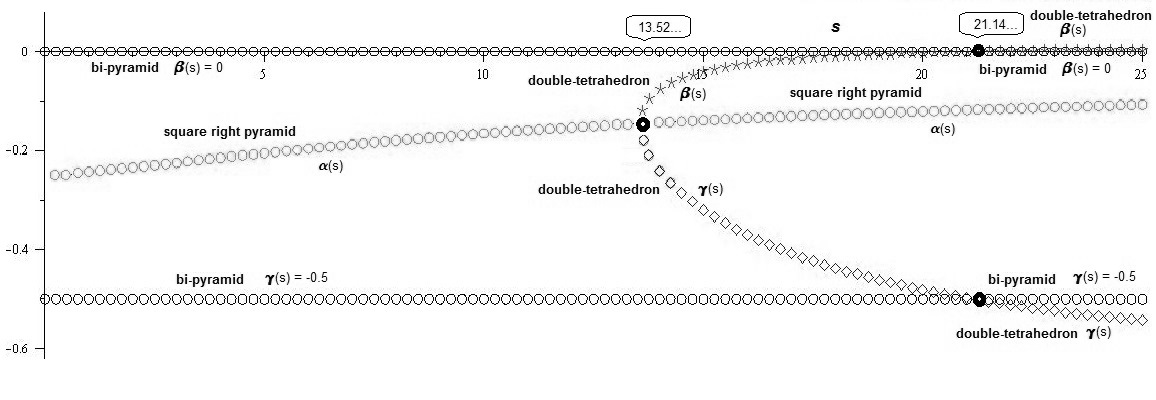}
  \caption{At vale $s=13.5204990011...$ two branches of double-tetrahedron configuration appear. These branches intersect the bi-pyramidal branches at value $s=21.1471229401...$ }\label{bifurcatie}
\end{figure}

Using MAPLE we find a third type of solutions for the system \eqref{eq-beta-gamma}, when $s>13.5204990011...$, that correspond to double-tetrahedron configuration. 
For $s\in (13.5204990011...,\, 21.1471229401...)$ we obtain solutions with the property $-1<\gamma<\beta<0$. For $s>21.1471229401...$ we obtain solutions with the property $-1<\gamma<0<\beta<1$. As predicted by Morse theory, the double-tetrahedron configuration has Morse index one (i.e. the Hessian matrix of $\wt{G}_s$ at this critical points has one strictly negative eigenvalue) at any $s>13.5204990011...$ value. In Figure \ref{bifurcatie}  we summarize the bifurcation phenomenon that appears in the branches of square right pyramid configuration and  bi-pyramid configuration. 

For $s$ between 15.048077392... and 21.1471229401... double tetrahedron configuration was first discovered in \cite{zhang-du}. In the same paper it was also noted that double-tetrahedron configuration exists for some values of $s$ smaller than 15.048077392... value. 

The value of the Riesz s-energy on this critical configuration is given by 
$$\wt{G}_s(\text{double-tetrahedron})=\frac{1}{(\sqrt{4-4\beta^2})^s}+\frac{1}{(\sqrt{4-4\gamma^2})^s}+\frac{2}{(\sqrt{2-2\beta})^s}+\frac{2}{(\sqrt{2-2\gamma})^s}+\frac{4}{(\sqrt{2-2\beta\gamma})^s},$$
 where $(\beta,\gamma)$ is the solution of the equation \eqref{eq-beta-gamma}.
\medskip

{\bf Comments}
\medskip

{\bf I.} In accordance with Conjecture 3.1. form \cite{cohn} we obtain that only bi-pyramidal configuration and square right pyramid configuration are local minima for Riesz s-energy.
\medskip

{\bf II.} The limit case when $s\rightarrow 0$ is the case of logarithmic interaction or the so called Whyte problem. The problem of 5 points on the sphere for this case has been rigorously studied in \cite{dragnev}.
\medskip

{\bf III.} The other limit case is when $s\rightarrow \infty$ and it is  known as the Tammes problem or the best packing problem on the sphere. An insightful connection between this problem and the N-vortex problem on the sphere has been studied in \cite{newton}.
\medskip

\begin{figure}[h!]
  \centering
    \includegraphics[width=0.9\textwidth]{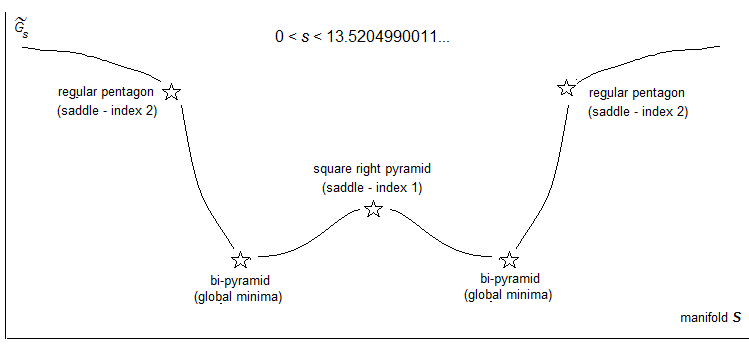}
  \caption{The distribution of the critical configurations are the same as in 5-electron Thomson problem.}\label{fox-1}
\end{figure}

\begin{figure}[h!]
  \centering
    \includegraphics[width=0.9\textwidth]{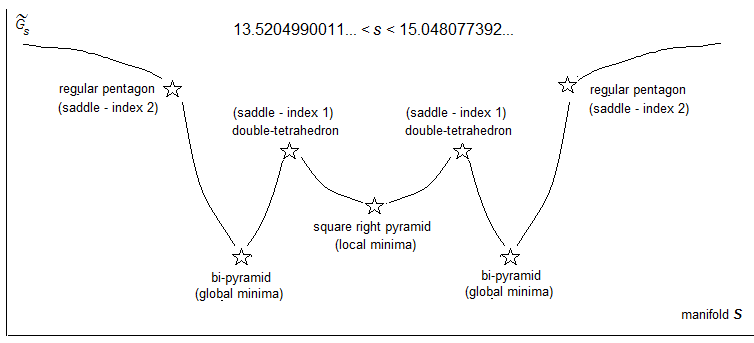}
  \caption{The first bifurcation behavior appears. The square right pyramid changes from saddle to local minima. Double-tetrahedron configuration appear as saddle critical points of Morse index one. }\label{fox-2}
\end{figure}

\begin{figure}[h!]
  \centering
    \includegraphics[width=0.9\textwidth]{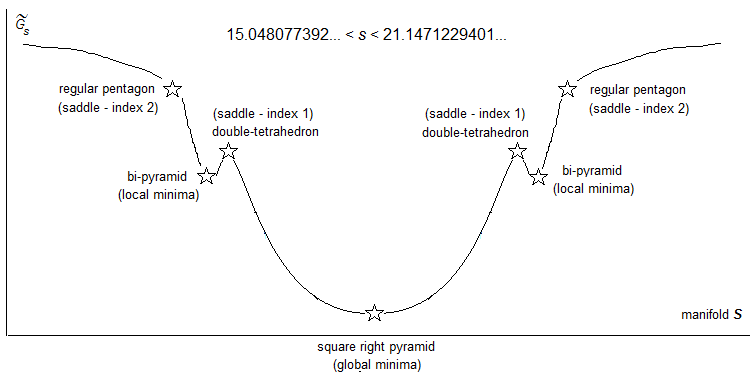}
  \caption{The square right pyramid configuration becomes global minima. }\label{fox-3}
\end{figure}

\begin{figure}[h!]
  \centering
    \includegraphics[width=0.9\textwidth]{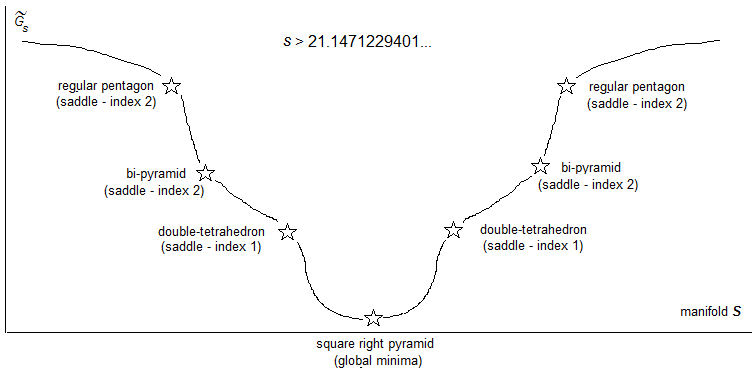}
  \caption{The bi-pyramid configuration becomes a configuration of saddle critical points with Morse index two.}\label{fox-4}
\end{figure}

{\bf IV.} In Figures \ref{fox-1}-\ref{fox-4} we draw the graphic of the cost function $\wt{G}_s$ on the eight dimensional manifold $S$, where we draw $S$ as one dimensional manifold. We also point out the bifurcation phenomenons that appears in Riesz s-energy problem.


\section{Appendix}

{\bf 1.} We give a short explanation for the equality $d\wt{G}(x)(\wt{{\bf w}}_{x})=dG(x)({\bf w}_{x})$. Around the point $x\in S$ we can find a system of local adapted coordinates $(x_i,y_{\alpha})_{i=\overline{1,s},\,\alpha=\overline{1,k}}$ such that the submanifold is locally described by the equations $S=\{y_{\alpha}=0\,|\,\alpha=\overline{1,k}\}$. The relation between $\wt{G}$ and its extension $G$ is given by $\wt{G}(x_i)=G(x_i,0)$. The differential forms are respectively
$$d\wt{G}=\sum_{i=1}^s\frac{\partial \wt{G} }{\partial x_i}dx^i,\,\,\,dG=\sum_{i=1}^s\frac{\partial G }{\partial x_i}dx^i+\sum_{\alpha=1}^k\frac{\partial G }{\partial y_{\alpha}}dy^{\alpha}.$$
The tangent vector $\wt{{\bf w}}_x\in T_xS$ can be written in the ambient coordinates as ${\bf w}_x=\sum_{i=1}^s w_i\frac{\partial}{\partial x_i}_{|x}$. Consequently, 
$$dG(x)( {\bf w}_x)=\sum_{i=1}^s\frac{\partial G }{\partial x_i}(x)w_i=\sum_{i=1}^s\frac{\partial \wt{G} }{\partial x_i}(x)w_i=d\wt{G}(x)(\wt{{\bf w}}_{x}).$$
\medskip

{\bf 2.} The Lagrange multiplier functions $\sigma_{\alpha}:M\rightarrow \R$, for $\alpha=\overline{1,k}$, are given by, see \cite{birtea-comanescu} and \cite{Birtea-Comanescu-Hessian}
\begin{equation}\label{sigma}
\sigma_{\alpha}(x):=\frac{\det \Sigma_{({F}_{1},\ldots ,{F}_{\alpha-1},{G}, {F}_{\alpha+1},\dots,{F}_{k})}^{({F}_{1},\ldots ,{F}_{\alpha-1},F_{\alpha}, {F}_{\alpha+1},\dots,{F}_{k})}(x)}{\det \Sigma_{({F}_{1},\dots,{F}_{k})}^{({F}_{1},\dots,{F}_{k})}(x)},
\end{equation}
with the Gramian matrix defined by
\begin{equation}\label{}
\Sigma_{(g_1,...,g_s)}^{(f_1,...,f_r)}=\left[%
\begin{array}{cccc}
  <\text{\bf grad } g_1,\text{\bf grad } f_{1}> & ... & <\text{\bf grad } g_s,\text{\bf grad } f_{1}> \\
  ... & ... & ... \\

  <\text{\bf grad } g_1,\text{\bf grad } f_r> & ... & <\text{\bf grad } g_s,\text{\bf grad } f_r> \\
\end{array}%
\right],
\end{equation}
where $<\cdot,\cdot>$ is the metric on the ambient space $M$.
\medskip

{\bf 3.} We construct the stereographic local frame on the sphere $S^2\subset \R^3$. The stereographic local chart from the North pole is given by: $\Psi:S^2\backslash \{\text{North Pole}\}\rightarrow \R^2$, 
$$\Psi(x,y,z)=(\xi_1,\xi_2)=(\frac{x}{1-z},\frac{y}{1-z}).$$
The inverse of this local chart is given by: $\Psi^{-1}:\R^2\rightarrow S^2\backslash \{\text{North Pole}\}$, 
$$\Psi^{-1}(\xi_1,\xi_2)=\left(\frac{2\xi_1}{1+\xi_1^2+\xi_2^2},\frac{2\xi_2}{1+\xi_1^2+\xi_2^2},\frac{-1+\xi_1^2+\xi_2^2}{1+\xi_1^2+\xi_2^2}\right).$$
The stereographic local frame is given by:
 $${\bf e}_1(x,y,z):=d\Psi^{-1}(\Psi(x,y,z))\cdot (1,0)=(1-z-x^2,-xy,x(1-z)),$$
 $${\bf e}_2(x,y,z):=d\Psi^{-1}(\Psi(x,y,z))\cdot (0,1)=(-xy,1-z-y^2,y(1-z)).$$


\begin{thebibliography}{99}


\bibitem{absil-book} {\bf P.A. Absil, R. Mahony, R. Sepulchre}, {\it Optimization Algorithms on Matrix Manifolds}, Princeton University Press, 2008.
\bibitem{ambrosetti} {\bf A. Ambrosetti, A. Malchiodi}, {\it Perturbation Methods and Semilinear Elliptic Problems on $\R^n$}, Birkh$\ddot{\text{a}}$user, 2006.
\bibitem{rabinowitz}{\bf A. Ambrosetti, P.H. Rabinowitz}, {\it Dual variational methods in critical point
theory and applications}, Jour. Funct. Anal., Vol. 14,  (1973), pp. 349-381.
\bibitem{ball}{\bf B. Ballinger, G. Blekherman, H. Cohn, N. Giansiracusa, E. Kelly, A. Schuermann}, {\it Experimental study of energy-minimizing point configurations on spheres}, Experimental Mathematics, Vol. 18 (2009), pp. 257-283.
\bibitem{birtea-comanescu} {\bf P. Birtea, D. Com\u anescu}, {\it Geometric Dissipation for dynamical systems},
    Comm. Math. Phys., Vol. 316, Issue 2 (2012), pp. 375-394.
\bibitem{Birtea-Comanescu-Hessian}{\bf  P. Birtea, D. Com\u anescu}, {\it Hessian Operators on Constraint Manifolds}, J. Nonlinear Science, Vol. 25, Issue 6 (2015), pp 1285-1305.
\bibitem{birtea-comanescu-popa} {\bf P. Birtea, D. Com\u anescu, C.A. Popa }, {\it Averaging on Manifolds by Embedding Algorithm},
    J. Math. Imaging Vis., Vol. 49, Issue 2 (2014), pp. 454-466.
\bibitem{bondarenko}{\bf A. V. Bondarenko, D. P. Hardin, E. B. Saff}, {\it Mesh ratios for
best-packing and limits of minimal energy configurations}, Acta
Math. Hungarica, Vol. 142, Issue 1 (2014), pp. 118-131.
\bibitem{bott}{\bf R. Bott}, {\it Nondegenerate critical manifolds}, Ann. of Math., Vol. 60, (1954), pp. 248-261.
\bibitem{cohn}{\bf H. Cohn}, {\it Order and disorder in energy minimization}, Proceedings of
the international congress of mathematicians (ICM 2010), pp. 2416-2443, 2010.
\bibitem{cohn-kumar}{\bf H. Cohn, A. Kumar},{\it Universally optimal distribution of points on spheres}, Journal of the American Mathematical Society, Vol. 20 (2007), pp. 99-148.
\bibitem{dragnev}{\bf P.D. Dragnev, D.A. Legg, D.W. Townsend}, {\it Discrete logarithmic energy on the sphere}, Pacific Journal of Math., Vol. 207, Issue 2 (2002), pp. 345-358.
\bibitem{gillespie}{\bf R.J. Gillespie}, {\it Fifty years of the VSEPR model}, Coordination Chemistry Reviews, Vol. 252, Issue 12-14 (2008), pp. 1315-1327. 
\bibitem{gillespie-hargitai}{\bf R.J. Gillespie, I. Hargittai}, {\it The VSEPR Model of Molecular Geometry},
Allyn and Bacon, Boston, 1991.
\bibitem{hargitai-chamberland}{\bf I. Hargittai, B. Chamberland}, {\it The VSEPR Model of Molecular Geometry}, Comp. $\&$ Maths. with Appls., Vol. 12B (1986), pp. 1021-1038.
\bibitem{huo-shao}{\bf Xiaorong Huo, Junwei Shao}, {\it Spherical Distribution of 5 Points with Maximal Distance Sum}, Discrete Comput Geom, Vol. 46, (2011), pp. 156-174.
\bibitem{melnyk}{\bf T. W. Melnyk, O. Knop, W. R. Smith}, {\it Extremal arrangements of points and unit charges
on a sphere: equilibrium configurations revisited}, Can. J. Chem., Vol. 55, (1976),  pp. 1745-1761.
\bibitem{muller}{\bf T. M\"{u}ller, J. Frauendiener}, {\it Charged particles constrained to a curved surface}, European Journal of Physics, Vol. 34, Issue 1 (2013), pp. 147-160. 
\bibitem{newton}{\bf P.K. Newton, T. Sakajo}, {\it Point vortex equilibria and optimal packings of circles on a sphere}, Proc. R. Soc. A, Vol. 467 (2011), pp. 1468-1490.
\bibitem{nicolaescu}{\bf L. Nicolaescu}, {\it An Invitation to Morse Theory}, Universitext, Springer, Second Edition 2011.
\bibitem{schwartz}{\bf R. E. Schwartz}, {\it The Five-Electron Case of Thomson's Problem}, Experimental Mathematics, Vol. 22, Issue 2, (2013), pp. 157-186.
\bibitem{zhang-du}{\bf J. Zhang, Q. Du}, {\it Constrained shrinking dimer dynamics for saddle point search with
constraints}, J. Comput. Phys., Vol. 231, (2012), pp. 4745-4758.



\end{thebibliography}
\end{document}